\documentstyle[preprint,aps]{revtex}
\tighten

\begin{document}
\preprint{UCONN-95-06}
\title{Probing Nucleon Structure via High Energy Elastic
Scattering$^{\dagger}$}
\author{M.M. Islam*}
\address{Department of Physics, University of Connecticut, Storrs, CT 06269,
U.S.A.}
\maketitle
\normalsize
\begin{abstract}

Analyses of high energy elastic $pp$ and $\bar pp$ scattering data from
CERN ISR and SPS Collider seem to provide strong evidence in favor of the
gauged nonlinear $\sigma$-model of the nucleon. This model describes the
nucleon as a topological soliton and introduces the vector mesons $\omega
,\rho ,a_1$ as gauge bosons. The model, however, needs to be extended to
include an explicit quark sector, where left and right quarks interact via a
scalar field. A critical behavior of the scalar field results in a phase
transition to a condensed quark-antiquark ground state. The latter can
provide the outer cloud of the nucleon, which is responsible for diffraction
scattering. If the nucleon is probed deeper via high energy elastic
scattering, then evidence for the phase transition may emerge from a rapid
change in the behavior of $d\sigma/dt$.
\end{abstract}
\thispagestyle{empty}
\vspace*{9.0cm}
\line(1,0){430}

$^{\dagger}$Talk given at the VI$^{th}$ Blois Workshop on Elastic and
Diffractive Scattering (Blois, France, June 1995).

$^{*}$ e-mail: islam@main.phys.uconn.edu

\newpage

Following the birth of the quark model in the mid-sixties,
physicists have devoted a great deal of effort to determine the composite
structure of the nucleon. This has not led to a unique description of the
nucleon. Instead, it has generated a wide array of models, such as:
constituent quark model, MIT bag model, topological soliton model,
non-topological soliton model, color dielectric model, etc. The reason for
such a proliferation of nucleon models is not hard to see. Low energy
properties of the nucleon and low energy nucleon-nucleon interaction have
not been able to single out one model as incorporating the key features.
What I am going to present is that, surprisingly, there seems to be strong
evidence in favor of one of the models --- namely, the gauged nonlinear $%
\sigma $-model that describes the nucleon as a topological soliton and
introduces the vector mesons $\omega ,\rho ,a_1$ as gauge bosons$^{1)}$. The
model,
of course, needs to be extended to include explicitly a quark sector, where
left and right quarks interact via a scalar field and form a condensed $q\bar
q$ ground state ( a chiral condensate). The nucleon emerges as a topological
soliton embedded in this condensate.

Let us first see how high energy elastic $pp$, $\bar pp$
scattering gets linked to the low energy soliton model in our work. High
energy elastic $pp$ and $\bar pp$ scattering have been measured at the CERN\
ISR\ and SPS Collider over a wide range of energy: $%
\sqrt{s}=23-630\,$ GeV. These data have been analyzed by me
and my collaborators over a number of years$^{2-4)}$. From our
analyses, we arrived at the following phenomenological description. The
nucleon has an inner core and an outer cloud. Elastic scattering at high
energy is primarily due to two processes: (1) a glancing collision in which
the outer cloud of one nucleon interacts with that of the other giving rise
to diffraction scattering; (2) a hard collision in which one nucleon core
scatters off the other nucleon core via vector meson $\omega $ exchange,
while their outer clouds overlap and interact independently. In the small
momentum transfer region, diffraction dominates. As the momentum transfer
increases, the hard scattering takes over. In Fig.1, we show our fit to the
ISR elastic $pp$ data at $\sqrt{s}=53\,$GeV$^{4)}$. Solid
curve represents the calculated differential cross section. The dot-dashed
curve represents the differential cross section due to diffraction alone,
and the dashed curve represents the differential cross section due to hard
scattering alone. As can be seen, diffraction dominates in the forward
direction, while the hard scattering takes over as $|t|$ increases.
The fits we obtained showed that our phenomenological
model of the nucleon does provide a satisfactory description of high energy
elastic scattering. Furthermore, questions such as $\omega $
behaving as a spin-1 boson in our analyses can be answered as due to the
topological nature of the baryonic charge$^{5)}$.

Despite these encouraging developments{\normalsize , we face a major
problem at this point. Several groups have found that the gauged nonlinear $%
\sigma $-model, even though quite successful in describing the low energy
properties of the nucleon, consistently predicts too large a soliton mass ($%
\sim 1500\,$MeV) compared to the actual mass of the nucleon ( $939\,$MeV )}$%
^{6)}${\normalsize . To address this question, let us consider the
Lagrangian density of the Gell-Mann-Levy $\sigma $-model :
\begin{eqnarray}
{\cal L} &=&\bar \psi i\gamma ^\mu \partial _\mu \psi +\frac 12\left(
\partial _\mu \sigma \,\partial ^\mu \sigma +\partial _\mu {\bf \vec \pi
\cdot }\partial ^\mu {\bf \vec \pi }\right) -G\bar \psi \left( \sigma
+i\gamma ^5{\bf \vec \tau }\cdot {\bf \vec \pi }\right) \psi   \nonumber \\
&&-\lambda \left( \sigma ^2+{\bf \vec \pi }^2-f_\pi ^2\right) ^2,
\label{gml}
\end{eqnarray}
Let us write, $\sigma +i{\bf \vec \tau }\cdot {\bf \vec \pi =}$ $\zeta
\left( x\right) \,U\left( x\right) $, where $U\left( x\right)
=\text{exp}\left[i
{\bf \vec \tau \cdot \vec \varphi }\left( x\right)/ f_\pi\right] $, ${\bf \vec
\varphi }\left( x\right) $ is the Goldstone pion field and $f_\pi $is the
pion decay constant ; $\zeta \left( x\right) $ is a scalar field which
corresponds to the magnitude of $\left( \sigma ,{\bf \vec \pi }\right) $
field: $\zeta ^2\left( x\right) =\sigma ^2\left( x\right) +{\bf \vec \pi }%
^2\left( x\right) .$ We can express the Lagrangian density (\ref{gml}) in
terms of the left and right fields $\psi _L=\frac 12\left( 1-\gamma ^5\right)
\psi $ and $\psi _R=\frac 12%
\left( 1+\gamma ^5\right) \psi :$
\begin{eqnarray}
{\cal L} &=&\bar \psi _Li\gamma ^\mu \partial _\mu \psi _L+\bar \psi
_Ri\gamma ^\mu \partial _\mu \psi _R+\frac 14\zeta ^2tr[\partial _\mu
U\partial ^\mu U^{\dagger }]+\frac 12\partial _\mu \zeta \partial ^\mu \zeta
\nonumber
\label{gml L2} \\
&&-G\zeta \left[ \bar \psi _LU\psi _R+\bar \psi _RU^{\dagger}\psi _L\right]
-\lambda
\left( \zeta ^2-f_\pi ^2\right) ^2.  \label{gml L2}
\end{eqnarray}
This model can now be easily gauged by
replacing the ordinary derivatives by the covariant derivatives. Using the
path-integral formalism, one then brings in a new piece of action---the
Wess-Zumino-Witten action. Two additional assumptions are made in the gauged
nonlinear $\sigma $-model : (1) all the important interactions are in the
meson sector (that is, among ${\bf \pi },\omega $, $\rho $ and $a_1)$, and
not in the fermion sector ; (2) the scalar field $\zeta \left( x\right) $
can be replaced by its vacuum value $f_\pi $ from the very beginning (this
makes the model nonlinear). Consequently, what one really has in the
nonlinear $\sigma $-model, is a soliton surrounded by a noninteracting Dirac
sea [Fig.2a]. If, instead, we consider the linear $\sigma $-model, then the
scalar field $\zeta \left( x\right) $ provides an interaction between the left
and right quarks, and we have an interacting Dirac sea surrounding the
soliton [Fig.2b]. One finds that, if the scalar field has a critical
behavior, and by this, I mean, it is zero at small distances, but rises
sharply at some distance $r=R$ to its vacuum value $f_\pi $ [Fig.2c], then
the energy of the interacting Dirac sea can be significantly less than that
of the noninteracting Dirac sea. The system, in that case, makes a phase
transition to the interacting ground state and reduces its total energy by
the condensation energy}$^{7)}${\normalsize . This can solve the large
soliton mass problem of the $\sigma $-model. Furthermore, the condensed
ground state of the left and right quarks is analogous to a superconducting
ground state. A number of important consequences follow from this result}$%
^{8)}${\normalsize . One of them is that the condensed $q\bar q$ ground
state can provide the outer cloud of the nucleon that we have in the
phenomenological description of elastic scattering. }

{\normalsize We may now ask : Is there any indication of the phase
transition considered above. The answer is : perhaps. Let us examine this
point. In elastic scattering, when the momentum transfer is $Q$, we are
probing the interaction at an impact parameter, i.e., at a transverse
distance of the order of $Q^{-1}$. Let us say the critical distance $R$ in
Fig.2c is $0.07\,F=(3\,$GeV$)^{-1}.$ (The reason for choosing this
particular value is pointed out later.) If we now consider elastic
scattering with $Q>3\,$GeV , then each nucleon probes the other nucleon at
an impact parameter less than $R$, i.e., in a region where the scalar field $%
\zeta \left( r\right) $ is zero, the quarks are massless, and we are in the
perturbative QCD regime. Elastic scattering in perturbative QCD has been
studied by Sotiropoulos and Sterman}$^{9,10)}${\normalsize . The scattering
originates from a valence quark in one nucleon scattering off a valence
quark in the other nucleon, and the differential cross section behaves as
\begin{equation}
\frac{d\sigma }{dt}\sim \frac{d\sigma _{qq}}{dt}\left( \frac{1/Q^2}{%
R_p^2}\right) ^8.  \label{x-section}
\end{equation}
The last factor in (\ref{x-section}) arises from the requirement that each
spectator valence quark has to be within a transverse distance of $1/Q$ of the
colliding valence quark}$^{9)}${\normalsize . The valence
quark-quark scattering in (\ref{x-section}) is assumed to be due to a color
singlet amplitude behaving as $t^{-1}$, so that $d\sigma _{qq}/dt%
\sim t^{-2}$ }{\normalsize . For $|t|>9$ GeV$^2$, this leads to the
power fall-off behavior$^{10)}$
\begin{equation}
\frac{d\sigma }{dt}\sim \frac 1{t^{10}}\,,  \label{x-section2}
\end{equation}
which corresponds to the dimensional counting rule. }

{\normalsize The behavior (\ref{x-section2}) is very different from the
Orear fall-off that we obtain for $Q<3$ $\,$GeV $(|t|<9$ GeV$^2)$ due to $%
\omega $ exchange:
\begin{equation}
\frac{d\sigma }{dt}\sim e^{-a\sqrt{|t|}}.  \label{orear}
\end{equation}
The length scale $a$ is connected with the finite size of the solitons. The
behavior (\ref{orear}) originates from a region where the scalar field $%
\zeta \left( r\right) $ has a nonvanishing value $f_\pi $, the quarks and
antiquarks form a condensed ground state, and we are in a nonperturbative
regime. So a rapid change in the behavior of $d\sigma/dt$ from an
exponential fall-off to a power fall-off will signal a transition from a
nonperturbative phase, where $\zeta \left( r\right) =f_\pi$, to a
perturbative phase, where $\zeta \left( r\right) =0.$ Interestingly enough,
the ISR $\sqrt{s}=53\,$GeV data indicate a change in the behavior of the
differential cross section for $|t|\simeq 9\,$GeV$^2$. As can be seen from
Fig.1, the last three data points show a flattening of the differential
cross section. (Our earlier choice of $R=\left( 3\,\text{GeV}\right) ^{-1}$
is, in fact, based on this observation.) The above scenario also implies
that the valence quarks are contained in a small region of size $R$ at the
center of the nucleon, so that when $Q<R^{-1}\;\left( |t|<9\,\text{GeV}%
^2\right) $, we see soliton-soliton scattering, but when $Q>R^{-1}\;\left(
|t|>9\,\text{GeV}^2\right) $, the elastic scattering originates from valence
quark-quark scattering. }
{\normalsize We note that before the perturbative QCD behavior $t^{-10}$
sets in, there can be an intermediate behavior $\left(d\sigma/dt%
\sim t^{-8}\right) $ preceding it. The latter behavior occurs when the three
valence quarks from one nucleon independently scatter off the three valence
quarks from
the other nucleon---a mechanism known as the Landshoff mechanism}$^{11)}$%
{\normalsize . }

{\normalsize In conclusion, our investigation of high energy $pp$ and $\bar p%
p$ scattering strongly indicates that the nucleon is a soliton that arises
from the topological baryonic charge distribution$^{5)}$. The latter is probed
by $%
\omega $, which acts like a photon. The soliton is embedded
in a condensed ground state of chiral quarks and
antiquarks. A rapid change in the behavior of the elastic differential cross
section from an exponential fall-off to a power fall-off at a large value of $%
|t|$ $\left( \sim 9\,\text{GeV}^2\right) $ will signal a QCD chiral phase
transition from a nonperturbative to a perturbative regime. Precise
systematic measurements of the large $|t|$ $pp$ and $\bar pp$ elastic
differential cross sections at the Tevatron, RHIC}$^{12)}${\normalsize , and
LHC}$^{13)}${\normalsize \ may reveal such a change in the behavior of $%
d\sigma/dt.$ It seems very exciting that from high energy elastic
scattering, we may find evidence of a chiral phase transition and of valence
quarks being confined at the center of the nucleon in a small region of size
about one-tenth of a Fermi. }

This work is supported in part by the U.S. Dept. of Energy.

\begin{center}
\Large References
\end{center}
1. I. Zahed and G.E. Brown, {\em Phys. Reports} {\bf 142} (1986) 1. \\
2. G.W. Heines and M.M. Islam, {\em Nuovo Cimento} {\bf A 61} (1981) 149.\\
3. M.M. Islam, T. Fearnley, and J.P. Guillaud,{\em Nuovo Cimento} {\bf A 81}
(1984) 737.\\
4. M.M. Islam, V. Innocente, T. Fearnley, and G. Sanguinetti, {\em Europhys.
Lett.}\\ \hspace*{4.7mm} {\bf 4}
(1987) 189.\\
5. E. Witten, {\em Nucl. Phys.} {\bf B 223} (1983) 422,433. \\
6. See, for example, L. Zhang and N.C. Mukhopadhyay {\em Phys. Rev.} {\bf D 50}
(1994) 4668 \\
7. M.M. Islam, {\em Z. Phys.} {\bf C 53} (1992) 253.\\
8. M.M. Islam, in {\em Proceedings of the Quantum Infrared Workshop}, ed. by
H.M. Fried and B.Muller (World
Scientific, 1995), p. 401. \\
9. M.G. Sotiropoulos and G. Sterman, in Int. Conf. on Elastic
and Diffractive Scattering (edited by H.M. Fried, K. Kang, and C-I. Tan; World
Scientific, 1994), p. 322. \\
10. M.G. Sotiropoulos and G. Sterman, {\em Nucl. Phys.} {\bf B 425} (1994) 489.
\\
11. Landshoff originally assumed exchanges of three gluons. Sotiropoulos and
Sterman point out that this
process has to be three color singlet exchanges (Ref.10).\\
12. At present, there is an approved $pp$ elastic scattering experiment at
RHIC, which will measure
$d\sigma/dt$ in the region $|t|= 0-1.5$ GeV$^2$. See W.Guryn's contribution to
this workshop.\\
13. The possibility of doing $\bar{p}p$ elastic scattering at the LHC is
discussed by K. Eggert in Proceedings of
this workshop.
\vspace*{11cm}

Fig. 1. Solid curve is the calculated cross section (see text for details).
\end{document}